\documentstyle[12pt,epsf,epsfig,wrapfig]{article}
\begin{document}
\begin{center}
{\bfseries VECTOR MESON PHOTOPRODUCTION AND PROBLEM OF GAUGE
INVARIANCE} \vskip 5mm S.V.~Goloskokov$^{\dag}$,  \vskip 5mm
{\small {\it Bogoliubov Laboratory of Theoretical Physics, Joint
Institute for Nuclear Research,\\ Dubna 141980, Moscow region,
Russia }
\\
$\dag$ {\it E-mail: goloskkv@thsun1.jinr.ru }}
\end{center}
\vskip 5mm

\begin{abstract}
We discuss the problem of gauge invariance of the vector meson
photoproduction  at small $x$ within the two-gluon exchange model.
It is found that the gauge invariance is fulfilled if one includes
the graphs with higher Fock states in the meson wave function. The
obtained results are used to estimate the amplitudes with
longitudinal and transverse photon and vector meson polarization.
\end{abstract}

\vskip 8mm

Investigation of  vector meson photoproduction at small $x$ is a
problem of considerable interest.  We are interested in the low-
$x$ region where the predominant contribution is determined by the
two-gluon exchange and the vector meson is produced via the
photon-two-gluon fusion. The factorization of diffractive vector
meson production with longitudinally polarized photons into the
hard part and parton distribution was shown in \cite{coll}. Thus,
 such processes, can be an excellent tool to
study the generalized parton distribution \cite{rad96}. Moreover,
they should give  important information on the vector meson wave
function. The  spin-density matrix elements which were studied at
DESY (see \cite{zeus} and references therein) should be sensitive
to the vector meson wave function. To analyze spin effects in the
$\gamma^\star \to V$ transition, it is necessary to calculate the
amplitude with transverse polarization of a vector meson. For the
light meson production, this transition amplitude is not well
defined because of the present end-point singularities
\cite{mank00}. One of the possible ways to regularize these
end-point divergences  is to include  the transverse quark motion,
as it was done, e.g., in \cite{ginzb,ivanov,cud,nikol}.

Unfortunately,  such higher-twist effects can result in  loss of
the gauge invariance (GI) of the amplitude. In this report, we
 study the $\gamma^\star \to V$ transition amplitude for
different polarization of photon and vector meson at small $x$ and
check the GI of our results. The  vector meson production can be
described in terms of the kinematic variables which are
follows:
\begin{equation}\label{momen}
q^2= (L-L')^2=-Q^2,\;r_P^2=(P-P')^2=t, \; x_P=\frac{q \cdot
(P-P')}{q \cdot P},\; s=(q+P)^2,
\end{equation}
where $L, L'$ and $P, P'$ are the initial and final lepton and
proton momenta, respectively, $Q^2$ is the photon virtuality,
$r_P$ is the momentum carried by the two-gluons, $x_P$ is  part of
proton momentum carried by the two-gluon system and $s$ is the
photon-proton energy squared. The vector meson is produced by the
photon-two-gluon fusion and the momentum $V=(q+r_P)$ is on the
mass shell. The $x_P$ variable which is equivalent to skewedness
$\zeta$ is determined by
\begin{equation}
\label{x_P} x_P \sim \zeta \sim \frac{M_V^2+Q^2+|t|}{s}.
\end{equation}

Within the two-gluon exchange model we  calculate the $L \to L$,
$T \to T$ and $T \to L$ amplitudes which are of importance in
analyses of spin density matrix elements. In calculations the $k$-
dependent wave function \cite{koerner} is used
\begin{equation}\label{psi}
\hat \Psi_V= g[ (\,/\hspace{-2.2mm} V+M_V) \,/\hspace{-2.9mm}E_V
                 + \frac{2}{M_V}\, /\hspace{-2.2mm} V \,
                 /\hspace{-2.9mm}
                 E_V /\hspace{-2.9mm} K
         - \frac{2}{M_V} (\, /\hspace{-2.2mm} V - M_V) (E_V \cdot K)]
         \phi_V(k,\tau).
\end{equation}
Here $V$ is a vector meson momentum and $M_V$ is its mass, $E_V$
is a meson polarization vector and $K$ is a quark transverse
momentum. The first term in (\ref{psi})  represents the standard
wave function of the vector meson. The leading twist contribution
to the longitudinal vector meson polarization is determined by the
$M_V \,/\hspace{-2.9mm}E_V$ term in (\ref{psi}). The $k$-
dependent terms of the wave function are essential for the
transverse amplitude of the light mesons. The wave function
(\ref{psi}) has quite a general form and can reproduce results of
most models \cite{ivanov,cud,nikol}. The other model for the wave
function which has a structure similar  to (\ref{psi}) was
considered in \cite{golos00}. The GI of the vector meson
production amplitude was discussed in \cite{ginzb,hebeck}. It was
found that the $\gamma^\star \to V$ transition amplitude at zero
momentum transfer should  vanish as $l_\perp^2$ for $l_\perp^2 \to
0$, where $l_\perp$ is the transverse part of the gluon momentum.
The importance of the higher Fock states of the wave function in
GI of the vector meson production was shown in \cite{hebeck}.
These results were obtained in the two-gluon model exchange in the
Feynman gauge.

The leading term of the amplitude of diffractive vector meson
production is mainly imaginary.  The imaginary part of the
amplitude can be written as an integral over $z$ and  $k_\perp$.
The leading over $s$   term of the $\gamma^\star \to V$  amplitude
has the form
\begin{equation}\label{Agv}
T_{\lambda_V,\lambda_\gamma} = N\,\int dz \int d k_\perp^2
 {\frac {{\cal F}^g_\zeta(\zeta,t)\, \phi_V(z, k_\perp^2)
A^{l^2}_{\lambda_V,\lambda_\gamma}(z,k_\perp^2)}
 {\left ({k_\perp^2}+{\bar Q^2}\right )\left( k_\perp^2+|t|+\bar Q^2
\right)^2}},
\end{equation}
where  N is normalization, $\bar Q^2=m_q^2+z \bar z Q^2$, $\bar
z=1-z$ and $m_q$ is a quark mass. Generally, the numerator of the
hard scattering amplitude $A_{\lambda_V,\lambda_\gamma}$  can be
written as follows:
\begin{equation}\label{ampl}
A_{\lambda_V,\lambda_\gamma}=A^0_{\lambda_V,\lambda_\gamma}+
A^{l^2}_{\lambda_V,\lambda_\gamma}\,l_\perp^2.
\end{equation}
Only the second term in (\ref{ampl}) obeys the GI and appears in
 (\ref{Agv}). The imaginary part of the vector meson production
amplitude (\ref{Agv})  depends on the  generalized gluon
distribution ${\cal F}^g_\zeta(X=\zeta,..)$. It can be connected
with the unintegrated gluon distribution $G$ through the
integration over $l_\perp$
\begin{equation}\label{gd}
{\cal F}^g_{\zeta}(\zeta,t,k_\perp^2+\bar Q^2+|t|)=
\int^{l_\perp^2<k_\perp^2+\bar Q^2+|t|}_0 \frac{d^2l_\perp
(l_\perp^2) } {(l_\perp^2+\lambda^2)((\vec l_\perp+\vec
r_\perp)^2+\lambda^2)} G(l_\perp^2,\zeta,...).
\end{equation}
Here $r_\perp$ is the transverse part of the $r_P$ momentum,
$\lambda$ is some effective gluon mass. The distribution ${\cal
F}^g_{0}(x,0,q_0^2)$ is normalized to $(x g(x, q_0^2))$. The
$l_\perp^2$ factor in the numerator of (\ref{gd}) appears from the
second GI term of (\ref{ampl}).

Unfortunately,  in the model with the higher twist effects like
 the transverse quark motion, the sum of  the graphs where
gluons are coupled with the quarks in the loop does not obey GI.
Let us discuss this problem in detail for the $L \to L$ amplitude.
The GI term of the amplitude has the form
\begin{equation}\label{al}
A^{l^2}_{L,L}=4 \frac{s}{\sqrt{Q^2}} \left[\bar Q^2 + k_\perp^2
(1-4 z \bar z) - 2 m_q M_V z \bar z \right] \,\left(\bar Q^2 +
k_\perp^2 \right).
\end{equation}
For the gauge-dependent term (GDT) we have
\begin{equation}\label{a0}
A^{0}_{L,L}=2 \frac{s}{\sqrt{Q^2}} \left[ k_\perp^2 (1-4 z \bar z)
+ m_q\,(m_q-2 M_V z \bar z )\right] \,\left(\bar Q^2 + k_\perp^2
\right)^2.
\end{equation}
It can be seen that in the nonrelativistic limit $z=\bar z =1/2$,
$m_q=M_V/2$ the GDT $A^{0}_{L,L}$ is equal to zero. For light
quarks, when $m_q=0$, the $A^{0}_{L,L}$ term is equal to zero at
$k_\perp^2=0$. At the same time, the $A^{0}_{L,L}$ term has
additional power of $\left(\bar Q^2 + k_\perp^2 \right)$ that
compensates one propagator in (\ref{Agv}) with respect to the GI
term (\ref{al}). As a result, the GDT  $A^{0}_{L,L}$ of the
amplitude (\ref{Agv}) is similar to the contribution of the higher
Fock state. Really, here  one gluon is coupled directly to the
wave function and one quark propagator disappears. Let us suppose
that we can write the sum of GDT and the contribution of the
higher Fock state in the form
\begin{equation}\label{sum}
\tilde A_{L,L} \sim A^{0}_{L,L}+ B(z,k_\perp^2)\, \Phi_{q \bar q
g}(1+C \,\frac{l_\perp^2}{Q^2}).
\end{equation}
Here by the $C \,l_\perp^2/Q^2$ term in (\ref{sum}) we estimate
the higher twist contributions in the $q \bar q g$ term of the
wave function. Let us  suppose that the higher Fock term
$B(z,k_\perp^2)\, \Phi_{q \bar q g}$ compensates the $A^{0}_{L,L}$
term in (\ref{sum}). In this case, the contribution proportional
to $l_\perp^2$  in $\tilde A_{L,L}$ can be estimated as
\begin{equation}\label{ab}
\tilde A_{L,L} \sim -C \,\frac{l_\perp^2}{Q^2}\, A^{0}_{L,L}.
\end{equation}
One can see that this GDT will be suppressed with respect to the
GI contribution (\ref{al}) as a power of $Q^2$. Really,
\begin{equation}\label{ratl}
 \frac{\tilde A_{L,L}}{l_\perp^2 \, A^{l^2}_{L,L}} \propto \frac{m_q^2+k_\perp^2}{Q^2}
\end{equation}
and we have GI of the model at sufficiently high $Q^2$.

Similar calculations have been done for the amplitude with
transversely polarized photons and vector mesons. The GI term of
this amplitude has the form
\begin{equation}\label{att}
A^{l^2}_{T,T}\sim \frac{2 s \, }{M_V}
 \bar Q^2\left[k_\perp^2 (1+4 z \bar z) +M_V \left(2 M_V z
\bar z-m_q (1-4 z \bar z)\right) \right]\, (e^\gamma_\perp
e^V_\perp).
\end{equation}
For light meson production the resulting amplitude  is
proportional to $k_\perp^2$. For heavy mesons, the term
proportional to $M_V^2$ appears too. In the transverse case, the
GDT which does not vanish as $l_\perp^2$ in (\ref{ampl}) takes
place like for the longitudinal amplitude. If we suppose the same
compensation of GDT as in (\ref{sum}), we  find
\begin{equation}\label{ratt}
\frac{\tilde A_{T,T}}{l_\perp^2 \, A^{l^2}_{T,T}} \propto z \bar
z.
\end{equation}
This means that in the transverse case we do not find a $Q^2$
suppression of additional GDT, but we have only its numerical
suppression. Really, it can be seen that the $T_{T,T}$ amplitude
has additional divergence like $1/(z \bar z)$ with respect to the
$T_{L,L}$ amplitude (\ref{Agv}). In the $\tilde A_{T,T}$ GDT the
additional $z \bar z$ term in the numerator cancels this
divergence and leads to the numerical suppression of the GDT
contribution.

The GI term of the $T \to L$ transition amplitude is determined by
\begin{equation}\label{atl}
A^{l^2}_{L,T}\sim \frac{2 s \, }{M_V}
 \bar Q^2\left[2 M_V^2 z \bar z-k_\perp^2 (1-2 z ) \right]\,
\frac{(e^\gamma_\perp r_\perp)}{M_V}
\end{equation}
It can be found that in this case we  have the numerical
suppression of a possible GDT contribution like for the $TT$
amplitude (\ref{ratt}).

Thus, we have found that the GDT in the $\gamma^* \to V$
transition amplitudes are suppressed and one can use the GI terms
(\ref{al}), (\ref{att}) and (\ref{atl}) to calculate
spin-dependent amplitudes of the vector meson production. The
average momentum transfer, which is used in  (\ref{Agv},
\ref{atl}) is about $<|t|> \sim 0.13 \mbox{GeV}^2$ \cite{zeus}.
The corresponding amplitudes were calculated for the $k-$
dependent wave function (\ref{psi}) with the exponential form of
$\phi_V(z, k_\perp^2)$ \cite{dis03}
\begin{equation}
\label{wf} \phi_V(z, k_\perp^2) = H  \exp{(-\frac{k_\perp^2
b_V^2}{z \bar z})}.
\end{equation}
Here $H$ is a  normalization factor. Transverse momentum
integration of (\ref{wf}) leads  to the asymptotic form of a meson
distribution amplitude $\Phi_V^{AS}=6 z \bar{z}$. The model has
one parameter $b_V$ which determines the average value of
$k_\perp^2$ and provides the regularization of the integrals in
the end- point region. In our calculation, we use the value $b_V
\sim 0.65 \,GeV^{-1}$ which  leads to a reasonable description of
the $\sigma_L$ cross section for $\rho$ production \cite{dis03}.
Then the average $<k_\perp^2> \sim 0.6 \, GeV^{2}$.

\bigskip
\begin{minipage}{7.6cm}
\phantom{aa} \hspace{-.4cm}\epsfxsize=7.5cm\epsfysize=6cm
\centerline{\epsfbox{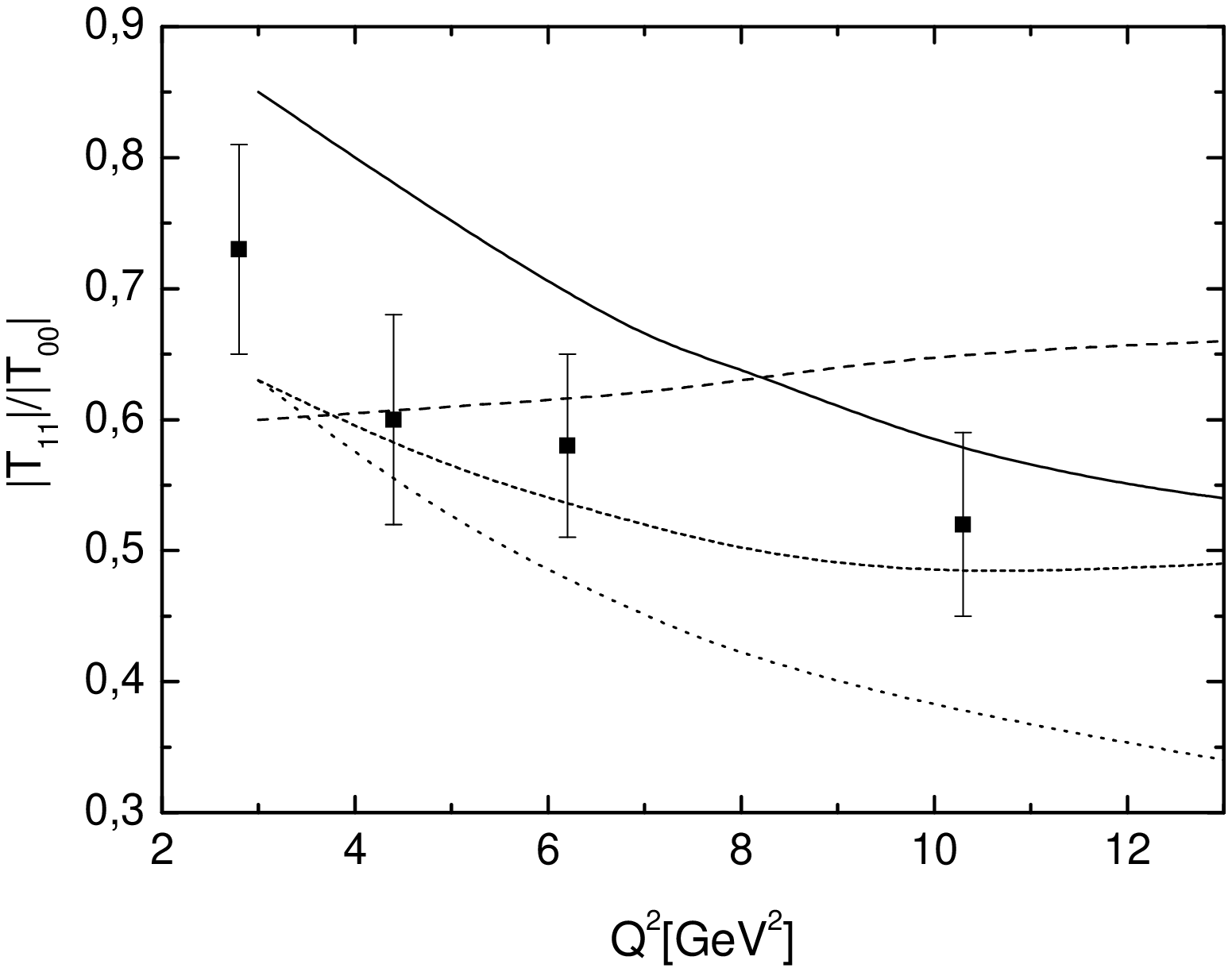}}
\end{minipage}
\begin{minipage}{0.01cm}
\phantom{aa}
\end{minipage}
\begin{minipage}{7.6cm}
\phantom{aa} \hspace{-.5cm} \epsfxsize=7.5cm \epsfysize=6cm
\centerline{\epsfbox{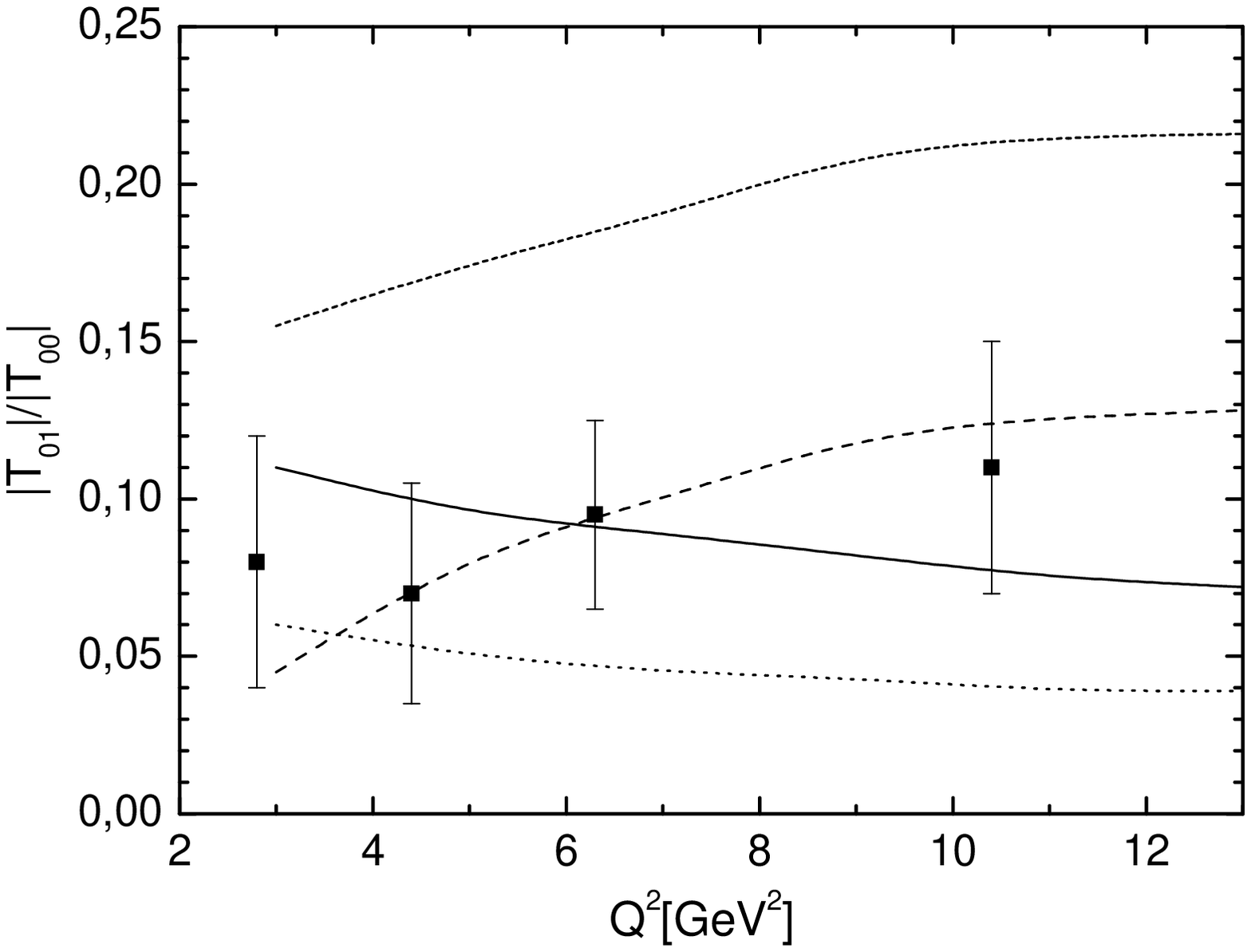}}
\end{minipage}
\\[3mm]
\begin{minipage}{7.6cm}
{\small{\bf Figure 1.} $Q^2$ dependence of the ratio of helicity
amplitudes $|T_{11}|/|T_{00}|$ \cite{zeus}. The full curve - our
calculation; dashed line -results of model \cite{ivanov}, short
dashed and dotted lines- models \cite{cud} and \cite{nikol}
respectively.}
\end{minipage}
\begin{minipage}{.6cm}
\phantom{aaa}
\end{minipage}
\begin{minipage}{7.5cm}
{\small{\bf Figure 2.} $Q^2$ dependence of the ratio of helicity
amplitudes $|T_{01}|/|T_{00}|$ extracted from H1 and ZEUS
measurements of the spin density matrix elements in \cite{zeus}.
Lines are the same as in Figure 1}.
\end{minipage}
\bigskip

The results of calculations for the  ratio of helicity amplitudes
$|T_{TT}|/|T_{LL}|$  are compared in Fig. 1 with the data
extracted in \cite{zeus} from H1 and ZEUS measurements of the spin
density matrix elements. It can be seen that experimental results
are reproduced by the model quite well. The model gives a
reasonable description of the ratio $R=\sigma_L/\sigma_T$. The
results of the models \cite{ivanov,cud,nikol} are shown in this
graph too. We can see that all the models describe experimental
data satisfactorily.

The comparison of model results for the $|T_{LT}|/|T_{LL}|$ with
experiment is presented in Fig.2. It can be seen that the $T_{LT}$
amplitude is more sensitive to the structure of the wave function.
The best description of the $|T_{LT}|/|T_{LL}|$ ratio  is found in
our model and in the models \cite{ivanov,nikol}. Note that the
experimental errors in the spin-density matrix elements are quite
large. This does not allow us to find out which model  of the wave
function describes experiment data adequately.

In this report, the results of the model for the $\gamma^* \to V$
transition amplitude which considers the transverse quark motion
have been analyzed. These higher twist effects regularize the
end-points singularities of the amplitudes but lead in the models
to violation of GI. Note that a similar problem with GI should
take place in the models \cite{ginzb, ivanov}. It is found that
the contribution of GDT in the model should be small. This permits
us to use  the model results for the GI terms of the $\gamma^* \to
V$ amplitudes for  numerical calculations. Our results describe
experimental data on the ratio of helicity amplitudes quite well.
Unfortunately, the experimental errors in DESY experiments are
large and all known models describe the experimental results
qualitatively.  To obtain more information on the form of the
vector meson wave function it is important to reduce the
experimental errors. We hope that the precise analyses of spin
density matrix elements can be done in the COMPASS experiment at
CERN.

This work was supported in part by the Russian Foundation for
Basic Research, Grant 03-02-16816.



\bigskip


\begin{thebibliography}{99}
\bibitem{coll}J.C. Collins, L. Frankfurt, M. Strikman,
Phys. Rev. D {\bfseries 56}, 2982 (1997).
\bibitem{rad96} D.\ M\"uller {\it et al},
                Fortschr. Physik {\bf 42}, 101 (1994);\\
                A.V.\ Radyushkin, Phys.\ Rev.\ D {\bf 56}, 5524
                (1997);\\
                X.\ Ji, Phys.\ Rev.\  D {\bf 55}, 7114 (1997).
\bibitem{zeus} B.\ Clerbaux for the H1 and ZEUS collaborations,
               hep-ph/9908519.
\bibitem{mank00} L.\ Mankiewicz, G. Piller, Phys. Rev. D {\bf 61}, 074013 (2000).
\bibitem{ginzb} I.F. Ginzburg, D.Yu. Ivanov, Phys.Rev. D {\bf 54}, 5523 (1996).
\bibitem{ivanov}
  D.Yu. Ivanov and R. Kirschner, Phys.Rev. D {\bf 59}, 114026 (1998).
\bibitem{cud}
 I. Royen and  J.-R. Cudell, Nucl. Phys. B {\bf 545}, 505 (1999).
\bibitem{nikol}
   I.P. Ivanov and N.N. Nikolaev, JETP Lett. {\bf 69}, 294 (1999); \\
   E.V. Kuraev, N.N. Nikolaev and B.G. Zakharov, JETP Lett. {\bf 68}, 696
   (1998).
\bibitem{koerner} J.~Bolz, J.~K\"orner and P.~Kroll,
Z.\ Phys.\ A {\bf 350}, 145 (1994).
\bibitem{golos00} S.V.\ Goloskokov, Proc. of XV International
  Seminar on High Energy Physics Problems "Relativistic Nuclear Physics and
  Quantum Chromodynamics", Dubna September 25-29, 2000,
  hep-ph/0012307.
\bibitem{hebeck} A.Hebecker, P.V. Landshoff,  Phys.\ Lett. B {\bf 419}, 393 (1998).
\bibitem{dis03} S.V.Goloskokov, P.Kroll, B.Postler, to appear in Proc. of
DIS03 workshop, St. Petersburg, 23-27 April 2003, hep-ph/0308140.
\end{thebibliography}
\end{document}